\documentclass[aps, prd, preprint, preprintnumbers, showpacs, amsfonts, nofootinbib, floatfix]{revtex4}
\usepackage{graphicx,epsfig}
\newcommand{\ba}{\begin{array}}
\newcommand{\ea}{\end{array}}
\newcommand{\bd}{\begin{displaymath}}
\newcommand{\ed}{\end{displaymath}}
\newcommand{\be}{\begin{equation}}
\newcommand{\ee}{\end{equation}}
\def\bt{\begin{table}}
\def\et{\end{table}}
\def\bc{\begin{center}}
\def\ec{\end{center}}
\def\bi{\begin{itemize}}
\def\ei{\end{itemize}}
\def\bea{\begin{eqnarray}}
\def\eea{\end{eqnarray}}
\def\beas{\begin{eqnarray*}}
\def\eeas{\end{eqnarray*}}

\def\gev{~\rm GeV}
\def\N0{\widetilde{\chi}^0}
\def\Dp{\widetilde{\Delta}^{++}}
\def\Dm{\widetilde{\Delta}^{--}}
\def\Dt{\widetilde{\Delta}}
\def\Cp{\widetilde{\chi}^+}
\def\Cm{\widetilde{\chi}^-}
\def\Cpm{\widetilde{\chi}^\pm}

\def\sel{\widetilde{e}_{L}}
\def\ser{\widetilde{e}_{R}}
\def\sml{\widetilde{\mu}_{L}}
\def\smr{\widetilde{\mu}_{R}}
\def\stl{\widetilde{\tau}_{1}}
\def\str{\widetilde{\tau}_{2}}

\def\D{\Delta}

\def\slash {\!\!\!\!/}

\catcode`@=11 
\def \gsim{\mathrel{\mathpalette\@versim>}}
\def \lsim{\mathrel{\mathpalette\@versim<}}
\def \@versim#1#2{\lower0.4ex\vbox{\baselineskip\z@skip\lineskip\z@skip
     \lineskiplimit\z@\ialign{$\m@th#1\hfil##\hfil$%
     \crcr#2\crcr\sim\crcr}}}
\catcode`@=12 
\begin{document}

\preprint{CUMQ/HEP 153, HIP-2009-06/TH, IZTECH-P-09/02}
\vspace*{0.01in}
\title{\Large  Doubly Charged Higgsinos at Tevatron}
\author{Durmu\c{s} A. Demir$^1$}
\email[]{demir@physics.iztech.edu.tr}
\author{Mariana Frank$^2$}
\email[]{mfrank@alcor.concordia.ca}
\author{Dilip Kumar Ghosh$^3$}
\email[]{dilipghoshjal@gmail.com}
\author{Katri Huitu$^4$}
\email[]{huitu@cc.helsinki.fi}
\author{Santosh Kumar Rai$^4$}
\email[]{santosh.rai@helsinki.fi}
\author{Ismail Turan$^2$}
\email[]{ituran@physics.concordia.ca}
\affiliation{$^{(1)}$Department of
Physics, Izmir Institute of Technology, IZTECH, TR35430 Izmir,
Turkey,}
\affiliation{$^{(2)}$Department of Physics, Concordia
University, 7141 Sherbrooke St.
 West, Montreal, Quebec, Canada H4B 1R6,}
\affiliation{$^{(3)}$Department of Theoretical Physics and Centre for Theoretical
Sciences, 
Indian Association for the Cultivation of Science,
2A \& 2B Raja S.C. Mullick Road, Kolkata 700 032, India}
\affiliation{$^{(4)}$Department of Physics, University of Helsinki,
and Helsinki Institute of Physics, P.O. Box 64,
     FIN-00014 University of Helsinki, Finland.}

%
\vspace*{0.2in}
\begin{abstract}
Several supersymmetric models with extended gauge structures,
motivated by either grand unification or by neutrino mass
generation, predict light doubly-charged Higgsinos. In this
work we study the signals of doubly-charged Higgsinos at the
Tevatron  in both pair-- and single--production modes, and show
that it is possible, especially from the events containing same-sign
same-flavor isolated leptons, to disentangle the
effects of doubly-charged Higgsinos in the Tevatron data.
\end{abstract}

\vspace*{-0.9cm}
\maketitle

\section{Introduction}
The long-awaited Large Hadron Collider (LHC), while delayed by
technical problems,  is expected to start soon. At the
beginning, it is very likely that the LHC will run at low beam
energy ($\approx 10$ TeV ) and low luminosity ($\approx
10^{32}~{\rm cm}^{-2} {\rm s}^{-1} $) regime. The data
collected during this period will be mainly used to calibrate
different components of the detectors involved in this
complicated experiment. In addition to the detector
calibration, the early measurements of $W$ and $Z$ production
cross sections will be able to provide a precise knowledge of
different parton densities at such high energies. While the LHC
may take at least one or two years to provide us the first
glimpse of the Higgs boson of the Standard Model (SM) or any
signature of new physics beyond the SM, it is worth
exploring the physics potential of the currently running
largest energy collider facility, Tevatron at Fermilab. It
should be noted that the Tevatron experiments D0$\,{\slash }$
and CDF have each recorded over $4~{\rm fb}^{-1}$ (Fall 2008)
data \cite{stefan}. Very recently, the combined analysis from
D0$\,{\slash }$ and CDF experiments based on data samples with
luminosities between $1.7-3$ $ {\rm fb}^{-1} $ excluded a SM
Higgs boson with a mass of 170 GeV at $95\%$ C.L.\cite{stefan}.
It has been speculated that the Tevatron will run through 2010
and at the current pace, it is aiming for over $8~{\rm
fb}^{-1}$ of data collection by these two experiments by the
end of 2010 \cite{stefan}. With these huge data sets, the
Tevatron will be able to probe the SM Higgs boson masses from
145 GeV to 185 GeV. In addition to Higgs search, one can
take advantage of this data set to explore
different scenarios of physics beyond the SM at the TeV scale.
The supersymmetric left-right model (which we hereon call
`LRSUSY') is one such example, which is based on the gauge
group $SU(3)_{\rm C} \times SU(2)_{\rm L} \times SU(2)_{\rm R}
\times U(1)_{{\rm B-L}} $. The LRSUSY model naturally arises
from high-scale models unifying left-- and right--handed matter, 
such as superstrings \cite{Frank_Sher} or supersymmetric GUTs
like $E_6$ or SO(10) \cite{Dutta_Mimura}. The LRSUSY model,
where the $SU(2)_{\rm R}$ gauge symmetry is broken by a triplet
Higgs field with quantum numbers ${\rm B -L }=\pm 2 $ has
several attractive features:
\begin{enumerate}
\item In LRSUSY,  $R$-parity ($R=(-1)^{(3{\rm B}+{\rm
    L}+2{\rm S})}$) is automatically conserved (with $B$,
    $L$ and $S$ baryon, lepton and spin quantum
    numbers, respectively), since the $B-L$ is part of the
    gauge symmetry \cite{RNMohaptra}. Therefore, the
    lightest supersymmetric particle (LSP) is automatically
    stable and it qualifies to be a candidate for cold dark
    matter \cite{Demir:2006ef}. Therefore events with
    supersymmetric particles are generically accompanied
    by  missing energy \cite{susysearch},  component
    taken away by the LSP.

\item In LRSUSY, strong and weak CP problems, the two
    serious naturalness problems which the minimal
    supersymmetric model (MSSM) suffers, are naturally
    avoided \cite{Mohapatra:1995xd}.

\item In LRSUSY, when $SU(2)_{\rm R}$ is broken by a
    triplet Higgs, the neutrino masses can be generated in
    a natural way by allowing for $R$-parity violating
    couplings through spontaneous $R$-parity violation, or by invoking the seesaw mechanism
    \cite{Mohapatra:1979ia}. The seesaw mechanism is
    natural in this model, where a Higgs triplet provides
    the Majorana mass term for the neutrinos (a detailed
    account of these aspects can be found in
    \cite{Mohapatra:1995xd,history,Francis:1990pi,Huitu:1993gf}).
\end{enumerate}
The Higgs triplets consist of doubly-charged, singly-charged, and
neutral fields. By supersymmetry, each superfield has
bosonic (Higgs bosons) and fermionic (Higgsinos) degrees of
freedom. The doubly-charged
fields cannot mix with fields belonging to other
electric charge sectors. The doubly-charged Higgsinos, as for
any fermionic component in a supermultiplet, cannot obtain soft
SUSY-breaking masses, and their masses thus originate solely
from the superpotential. Indeed, the superpotential involves
bilinears of the Higgs triplets, and the corresponding $\mu$
parameter (in the language of MSSM) generates the requisite
masses for doubly-charged Higgsinos. In general, there is no
telling of whether the  $\mu$ parameter lies at the electroweak
scale (similar to the $\mu$ problem of the MSSM); however, if
it does, then Higgsinos weigh within the reach of present
colliders \cite{Chacko:1997cm}. The $\mu$ parameter can 
be generated and stabilized at the weak scale though some
variation of the Giudice-Masiero mechanism \cite{gm} or some
dynamical stabilization mechanism utilizing an extra gauge
group, under which the charges of left-- and right--triplets do
not sum up to zero  
\cite{muprob}.

The collider signals of the  doubly-charged Higgs bosons of
LRSUSY model have been studied in the context of upcoming
collider experiments \cite{Huitu:1996su}. The signatures of the
fermionic partners of the doubly-charged Higgs bosons have been
studied at the LHC \cite{Huitu:1995bc,Demir:2008wt} and at a
linear collider \cite{Huitu:1993gf,Frank:2007nv}. Moreover,
signatures of such doubly-charged Higgsinos of the LRSUSY
model have been explored previously  \cite{dutta_patra_muller},  and in the framework of gauge mediated supersymmetry
breaking scenario, at the
Tevatron. In this paper, we consider
the pair and single production of doubly-charged Higgsinos at
the Tevatron and their subsequent decays which lead to
multilepton plus large missing energy signatures. We perform a 
background analysis, and show how the signal can be extracted.
We then discuss the discovery potential of the Tevatron, and
demonstrate that it can probe doubly-charged Higgsino masses up
to 200-300 GeV at the present integrated luminosity.

The rest of the paper is organized as follows. In Section II we
describe the salient features of the LRSUSY model. In Section
III we discuss the pair and single production of doubly-charged
Higgsinos at the Tevatron energies and their decay channels. In
Section IV we analyze thoroughly the signal and background
events and discuss the possible discovery limits for doubly-charged higgsinos using the multilepton
signature. Finally, our conclusions are given in Section V.

\section{Doubly Charged Higgsinos in the Left Right Supersymmetric Model}
\label{model}
In what follows we choose LRSUSY to describe the interactions
of the doubly-charged Higgsinos, although we expect our
analysis to be general enough to hold in generic models (such
as the $3-3-1$ model) which predict such exotic particles.
LRSUSY adds supersymmetry to the left-right gauge symmetry
group
\begin{eqnarray}
G_{LR} \equiv SU(2)_{L}\times SU(2)_{R}\times U(1)_{B-L}
\end{eqnarray}
in addition to color $SU(3)$. The matter spectrum consists of
three generations of quark and lepton superfields as well as
the gauge bosons of each group factor~\cite{Francis:1990pi}.
The Higgs sector is spanned by
\begin{itemize}
\item Two $B-L = 0$ Higgs bi-doublets: $\Phi_{1}(2,2,0)$
    and $\Phi_{2}(2,2,0)$ required to generate
    non-vanishing CKM quark mixing.
\item Two $B-L = -2$ Higgs triplets:  $\Delta_{L}(3,1,-2)$
    and $\Delta_{R}(1,3,-2)$. One is  required to break
    $SU(2)_{R}\times U(1)_{B-L}$ down to  $U(1)_{Y}$
    spontaneously and both are needed to preserve left-right symmetry.
\item Two $B-L = +2$  Higgs triplets: $\delta_{L}(3,1,2)$
    and $\delta_{R}(1,3,2)$ required to cancel the
    anomalies.
\end{itemize}
Given this superfield spectrum, the superpotential of LRSUSY
takes the form
\begin{eqnarray}
\label{superpotential}
W & = & {\bf Y}_{Q}^{(i)} Q^T\tau_{2}\Phi_{i} \tau_{2}Q^{c} + {\bf Y}_{L}^{(i)}
L^T\tau_{2}\Phi_{i} \tau_{2}L^{c} + i {\bf Y}_{LL}\left[L^T\tau_{2} \delta_L L +
L^{cT}\tau_{2} \Delta_R  L^{c}\right] \nonumber \\
& + & \mu_3 \mbox{Tr}\left[ \Delta_L  \delta_L + \Delta_R
\delta_R \right] + \mu^{(ij)} \mbox{Tr}\left[\tau_{2}\Phi^{T}_{i} \tau_{2} \Phi_{j}\right] + W_{NR}
\end{eqnarray}
where generation indices are suppressed. $W_{NR}$ stands for
possible non-renormalizable operators. The dimensionful
parameters $\mu$ and $\mu^{(i j)}$ will be taken to have been
stabilized at the electroweak scale without specifying the
responsible mechanism \cite{gm,muprob}. The ${\bf Y}_Q^{(i)}$
and ${\bf Y}_L^{(i)}$ are, respectively, the Yukawa matrices
for quarks and leptons for their couplings to the bi-doublet
$\Phi_{i}$. The ${\bf Y}_{LL}$ parameterizes couplings of left-- and
right--handed leptons to the corresponding Higgs triplet. The
left-right symmetry enforces ${\bf Y}_{Q, L}$ matrices to be
Hermitean and ${\bf Y}_{LL}$ matrix to be symmetric in the
space of fermion generations.

The parity and $SU(2)_{\rm R}$ symmetries are spontaneously
broken by the non-vanishing VEVs of the Higgs mutiplets
\begin{eqnarray}
\langle \Delta_{L,R} \rangle =\left(\begin{array}{cc}
0&v_{\Delta_{L,R}}\\ 0&0
\end{array}\right)\,,
\nonumber
~~~\langle \delta_{L,R} \rangle  = \left(\begin{array}{cc}
0& 0\\ v_{\delta_{L,R}}&0
\end{array}\right)\,,\;\;
\langle \Phi_{1,2} \rangle = \left (\begin{array}{cc}
v_{1,2}&0\\0& v^{\prime}_{1,2}
\end{array}\right)
\end{eqnarray}
where position of various non-vanishing entries are dictated
by the electric charge conservation. Clearly, the VEVs of the
bi-doublets $v_{1,2}$ and $v^{\prime}_{1,2}$ are responsible
for giving mass to quarks and leptons. The VEV $v_{\delta_L}$
must be exceedingly small for neutrino masses to be generated
correctly (if ${\bf Y}_{LL}$ is not taken unnaturally small)
and for the $\rho$ parameter to stay close to unity (which also
constrains $v_{\Delta_L}$ to be small). The VEV $v_{\Delta_R}$
must lie close to right-handed neutrino scale if neutrino
masses are to be generated by see-saw mechanism. If  $v_{\Delta_R}$ 
along  with $v_{\delta_R}$, are chosen to stay close to ${\rm TeV}$
domain if $W_R$ and $Z_R$ are to be seen at collider
experiments, one must seek alternative ways to generate neutrino masses.

Though $\Delta_{L}$ is not necessary for symmetry
breaking~\cite{Huitu:1996su} as it is introduced only for
preserving left-right symmetry, both $\Delta_{L}^{--}$ and its
right-handed counterpart $\Delta_{R}^{--}$  play very important
roles in phenomenological studies of the LRSUSY model. It has
been shown that these bosons, and  their fermionic
counterparts, could be sufficiently light \cite{Chacko:1997cm}
to be reachable in present colliders. As our analysis deals
primarily with the doubly charged Higgsinos, we hereon focus on
these doubly-charged states.

The mass terms of doubly-charged Higgsinos read as
\begin{equation}
{\cal L}^{mass}_{\Delta\delta}=-M_{\tilde \Delta_L^{--}}
                   {\tilde \Delta}_{L}^{--}{\tilde {\delta}}_L^{++}
                -M_{\tilde \Delta_R^{--}}
                   {\tilde \Delta}_{R}^{--}{\tilde {\delta}}_R^{++}
\end{equation}
where
\begin{eqnarray}
M_{\tilde \Delta_L^{--}} =M_{\tilde \Delta_R^{--}} \equiv \mu_3
\end{eqnarray}
as follows from the superpotential (\ref{superpotential}). This
relation can be further affected by the contributions of the
non-renormalizable operators in $W_{NR}$.

The Yukawa interactions of doubly-charged Higgsinos read as
\begin{equation}
{\cal L}_{Y}=-2\,{Y}^{ij}_{LL}{\bar L}^c_{iL}
               {\tilde \delta}_L^{++} { \tilde L}_{jL}
             -2\, {Y}^{ij}_{LL}{\bar L}^c_{iR}
               {\tilde {\Delta}}_R^{--}{\tilde L}_{jR}
\end{equation}
as follows from the superpotential (\ref{superpotential}). We
consider only the flavor-diagonal Yukawas ($i=j$), for
simplicity. The advantages of studying the production and
decays of the doubly-charged Higgsinos is evident. Their masses
and interactions do not depend on the parameters in other
gaugino and Higgsino sectors. In addition, ${\tilde
\Delta}_L^{--}$ and ${\tilde \Delta}_R^{--}$ do not mix with
each other, so their interactions depend on a small set of
unknowns. In fact, the only parameters are their masses
$M_{{\tilde \Delta}^{--}_L} = M_{{\tilde \Delta}^{--}_R}$ and
leptonic Yukawa couplings $Y^{ii}_{LL}$, which we denote by $Y_{LL}$ from here on. 
In what follows we forgo further details of the model and proceed directly to
analyze the production and decays of the doubly-charged
Higgsinos at the Tevatron. For additional details about the
model, including discussion of the singly-charged charginos and
neutralinos, as well as the scalar lepton sector, we refer the
reader to our previous work \cite{Demir:2008wt}.
\begin{center}
\bt[htb]
$$
\begin{array}{|c|c|}\hline
           & {\rm \bf SPA} \\
           & {\it M_{B-L}}= 25\gev, {\it M_{L}}={\it M_{R}}=250\gev \\
\mbox{Fields} &\tan\beta=5,v_{\D_R}=3000\gev, {\it v_{\delta_R}}=1000\gev \\
              &\mu^{(1 1)} = \mu^{(2 2)} = 1000\gev,\mu_3 = 200\gev \\
\hline \hline
\N0_i~(i=1,3) & 92.2,120.9,200\gev \\
\Cpm_i~(i=1,3)& 200,250.9,934.7\gev \\
  W_R, Z_R    & 2090.4, 3508.5\gev  \\
(\sel,\ser),(\sml,\smr),(\stl,\str) & (402,402\gev),(402,402\gev),(401,406\gev) \\ \hline \hline

\end{array}
$$
\caption{\sl\small The numerical values assigned to the model parameters in
defining the sample point {\bf SPA}. In the list, $M_{B-L}$, $M_{L}$ and $M_{R}$ are, respectively, the masses of
$U(1)_{\rm B-L}$, $SU(2)_{\rm L}$ and $SU(2)_{\rm R}$ gauginos.
The VEVs of the left-handed Higgs triplets are taken as
$v_{\D_L}\sim v_{\delta_L}\simeq 10^{-8}\gev$. For gauge
couplings we take $g_L=g_R=g$, and for Yukawas we take
$Y_{LL}=0.1$. Masses of some light eigenstates are shown in the Table.
} \label{susyin} \et
\end{center}

\section{Production and Decay of Doubly Charged Higgsinos at Tevatron}
\label{sec:results}
We focus now on the production modes for doubly-charged
Higgsinos at Tevatron. Single-- as well as pair--production of
the doubly-charged Higgsinos can take place through the
$s$-channel exchange of the relevant gauge bosons in the model.
The pair--production process at the Tevatron  $p\,
\bar {p} \longrightarrow \Dp\, \Dm $, proceeds with
$s$-channel $\gamma$ and $Z_{L,R}$ exchanges, and the single production, $p\, \bar {p}
\longrightarrow \Cp_1\, \Dm $, is dominated by
$s$-channel $W_{L,R}$ exchanges. Both processes are initiated
by quark--anti-quark annihilation at the parton level at the
Tevatron.

These doubly-- and singly--charged fermions subsequently decay
via a chain of cascades until the lightest neutralino
$\widetilde{\chi}_1^0$ is reached. In general, the two-body decay modes of
doubly-charged Higgsinos are given by \bi
\item $\Dm \longrightarrow \widetilde{\ell}^- ~\ell^-$,
\item $\Dm \longrightarrow \Delta^{--} ~\N0_i$,
\item $\Dm \longrightarrow \Cm_i ~\Delta^{-}$,
\item $\Dm \longrightarrow \Cm_i ~W^{-}$,
\ei
whose decay products further cascade into lower-mass daughter particles of
which leptons are of particular interest. Clearly, pair-produced
doubly-charged Higgsinos can lead to $4 \ell + E\slash_T$ final states, 
whereas singly-produced doubly charged Higgsinos can give rise to 
$3 \ell + E\slash_T$ signals.

The possibility of light observable doubly-charged Higgs bosons
has been explored extensively by both phenomenological analyzes
\cite{dcbtheory} and experimental investigations, and they
resulted in constraining the two-dimensional parameter space
spanned by the doubly-charged Higgs mass and the $\Delta L=2$
coupling \cite{dcbexperiment}. We hereon concentrate on
accessible decay channels devoid of Higgs bosons. 
Thus we assume that triplet Higgs bosons are heavier and degenerate
in mass, which renders them kinematically inaccessible for
decay modes of the relatively lighter doubly-charged Higgsinos.
For the
numerical estimates we have considered a representative
point in the LRSUSY parameter space, favorable to observing light doubly charged higgsinos, as tabulated in Table
\ref{susyin}. A quick look at the resulting mass spectrum for
the sparticles suggest that the chargino states as well as the
scalar leptons are also heavier than the doubly-charged
Higgsinos, and hence, the favorable decay channel for $\Dt$
(for relatively light Higgsinos) would be the 3-body decays,
which would proceed dominantly through off-shell sleptons: $\Dm
\to \widetilde{\ell}^{\star\, -} ~ \ell^- \to \ell^- \ell^-
\N0_i$, where $\N0_i$ is decided by the allowed kinematic phase
space. In addition, the doubly-charged Higgsinos can have a
3-body decay through the heavy off-shell charginos. We have
explicitly checked that the 3-body decay of the doubly-charged
Higgsinos through the heavy off-shell charginos or $W$ bosons
\begin{figure}[htb]
\begin{center}
\includegraphics[height=3.0in,width=3.0in]{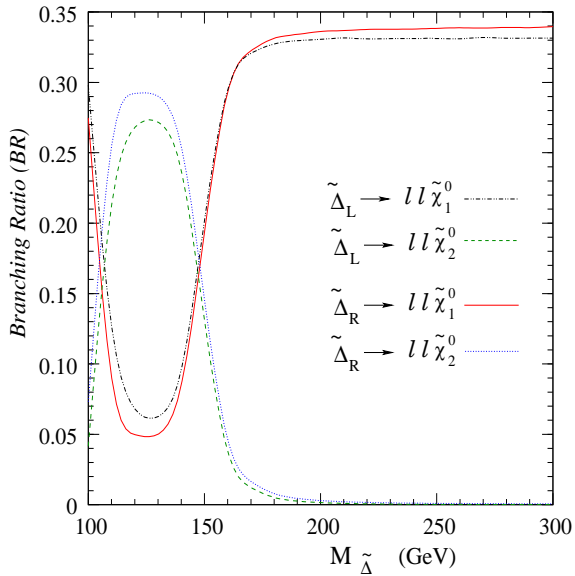}
\caption{\sl\small The 3-body decay branchings of $\Dm$
and $\Dp$ for the {\bf SPA} parameter set, but allowing $\mu_3$ to vary. The branching ratios change  mainly with what
neutralino, $\N0_1$ or $\N0_2$, is produced along with SSSF
lepton pairs ($\ell=e$ or $\mu$)}
\label{bratio}
\end{center}
\end{figure}
is quite suppressed (unless the $\Delta L=2$ coupling $Y_{LL}$
is fine-tuned to a very low value ($< 10^{-5}$)) with respect
to the 3-body decay through the off-shell sleptons, and can be
safely neglected. In Fig. \ref{bratio} we show the relative
branching ratios corresponding to the 3-body decays of light
doubly-charged Higgsinos as a function of their mass. We can
see that for a relatively light $\Dt$ of mass $M_{\Dt}< 150\gev$ the
mode $\ell\ell\N0_2$ is the dominant one due to the compositions
of light neutralinos, despite the fact that
${\widetilde \chi}_2^0$ is heavier than the LSP. Nevertheless,
it falls rapidly when $M_{\Dt} > 150\ {\rm GeV}$, and the mode
$\ell\ell\N0_1$ takes over. Consequently, one writes
\begin{eqnarray}
\label{eq:decays}
 BR(\Dm_{L/R} \to \ell_i^- \ell_i^-  \N0_1) &\simeq& \frac{1}{3},
~~~m_{\tilde l_{i}}>M_{\tilde\Delta^{--}}
\end{eqnarray}
where $i=e,\mu,\tau$. One notes that only the 3-body decay
channel is allowed when $m_{\tilde{\ell}_{i}}>M_{\Dt}$. We
discuss the chargino decays later when analyzing the single
production mode.

We now focus separately on the signals at Tevatron arising from the pair-- and
single--productions of these doubly-charged Higgsinos.

\subsection{Pair-production of doubly-charged Higgsinos}

The pair--production of doubly-charged Higgsinos at the
Tevatron occurs through the $s$-channel exchanges of the
neutral gauge bosons in the model {\it viz.} the $\gamma$,  $Z$
and the new (heavy) $Z$ boson ($Z_R$). Since $Z_R$ is heavy, it
has negligible contribution to the production at the Tevatron.
It is important to note that the production cross section of
the doubly-charged Higgsino which is a fermion, is much larger
than that of a doubly-charged Higgs of the same mass. The
production rate is mainly determined by the fact that the
photon couples to the two units of charge carried by the
doubly-charged Higgsino, and there is a kinematic suppression
in the phase space due to the mass of the doubly-charged state.
Therefore, we can safely assume that a single representative
point will be sufficient to present the essential features of
the signal one can expect at Tevatron.
\begin{figure}[ht]
\begin{center}
\includegraphics[height=3.0in,width=4.0in]{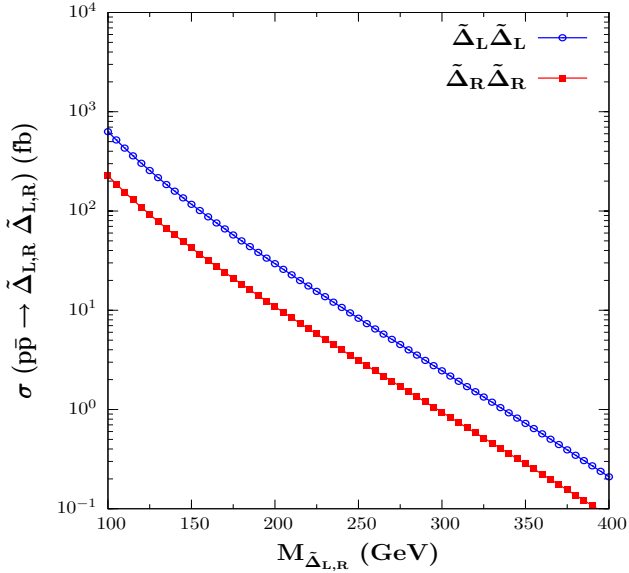}
\caption{\sl\small The pair-production cross sections for doubly-charged
Higgsinos of either chirality at the Tevatron. The plots are performed by using the
parameter sets {\bf SPA} except that $M_{\tilde \Delta^{--}} \equiv \mu_3$ is
allowed to vary from $100\ {\rm GeV}$ up to $400\gev$.}
\label{pairprod}
\end{center}
\end{figure}
In Fig.~\ref{pairprod} we plot production cross sections for
$\Dm$ for both chiralities and exchanged gauge bosons. It is
seen that the cross section is quite sizeable for sufficiently
light doubly-charged Higgsinos:  It starts at $\sim 10^{3}\
{\rm fb}$ at $M_{\Dt} \simeq 100\ {\rm GeV}$ and stays above
$\sim 10\ {\rm fb}$ for $M_{\Dt}$ up to $250\ {\rm GeV}$.
However, it falls rapidly below $\sim 1\ {\rm fb}$
 for masses beyond $M_{\Dt}=300\gev$.

The doubly-charged Higgsinos decay according to
(\ref{eq:decays}) into two same-sign same-flavor (SSSF) leptons
and the LSP $\N0_1$. This decay pattern gives rise to final
states involving four isolated leptons of the form
$\left(\ell_i^- \ell_i^-\right)\, \left(\ell_j^{+}
\ell_j^{+}\right)$:
\begin{equation}
p p \longrightarrow \Dp \Dm \longrightarrow  \left(\ell^+_i \ell^+_i\right) +
\left(\ell^-_j \ell^-_j\right) + E\slash_T\,,
\label{eq:pairprod}
\end{equation}
where $\ell_i$ and $\ell_j$ are not
necessarily identical lepton flavors, and where $\ell_i,\ell_j=e,\mu,\tau$.

The $4\ell+E\slash_T$ signal receives contributions from the
pair-production of both chiral states of the doubly-charged
Higgsino. Since at the Tevatron it is difficult to determine
chiralities of particles, it is necessary to add up their
individual contributions to obtain the total number of events.
This yields a rather clean and robust $4\ell+$ missing $E_T$
signal  with highly suppressed SM background. One
finds that the SM cross section with tetraleptons, where
$\ell_i=e$ and $\ell_j=\mu$,  in (\ref{eq:pairprod}) and with large
missing transverse energy ($E\slash_T \ge 25\gev$), receives
the dominant contribution from $t\bar{t}$ production, and, to a
lesser extent, from the pair-production of gauge bosons, $WW$
and $WZ$.

For the numerical analysis, we have included the LRSUSY model into
{\tt CalcHEP 2.4.5} \cite{calchep} and generated the event
files for the production and decays of the doubly-charged
Higgsinos using the {\tt CalcHEP} event generator. The event
files are then passed through the {\tt CalcHEP+Pythia}
interface, where we include the effects of both initial and
final state radiations using {\tt Pythia} \cite{pythia} switches
 to smear the final states. We use the leading
order CTEQ6L \cite{cteq} parton distribution functions (PDF)
for the (anti-)quarks  in (anti)-protons. We employ the
jet cone algorithm implemented in {\tt Pythia} through the
subroutine {\tt PYCELL}. We assume that the minimum summed
the jet transverse energy $E_T$  (consisting of all calorimetric cells within
the cone of radius $\Delta R=0.7$ in the ($\eta,\phi$) plane)
must be 15 GeV to qualify to be a jet. The final state leptons
are considered to be isolated if they are well resolved from
the jets by demanding $\Delta R_{\ell_i J} \ge 0.4$.

In addition, for triggering and enhancing the $4\ell+E\slash_T$ signal, we impose
the following kinematic cuts \cite{Abel:2000vs}:
\begin{itemize}
\item  The charged leptons in the final state must respect the rapidity cut
$|\eta_{\ell}|<2.0$,
\item  The charged leptons in the final state (arranged
    according to their $p_T$'s) must satisfy $p_T (\ell_1)
    > 11\gev$, $p_T (\ell_2) > 7\gev$ and $p_T (\ell_{3,4})
    > 5\gev$.
\item To ensure proper resolution between the final state
    leptons we demand $\D R_{\ell_i \ell_j}>0.2$ for each
    pair of leptons, where $ \D R = \sqrt{ (\D \phi)^2 +
    (\D \eta)^2 }$, $\phi$ being the azimuthal angle.
\item The missing transverse energy must obey $E\slash_T >
    25$ GeV.
\item The pairs of oppositely-charged leptons of the same
    flavor have at least $10\ {\rm GeV}$ invariant mass.
\end{itemize}

The SM background yielding tetralepton final states receives
the dominant contribution from the $t\bar{t}$ production,
which can be significantly suppressed by demanding at least two
same-sign muons and two same-sign electrons in the final state.
This removes the unwanted large contributions produced at the
$Z$-peak. We simulate the SM background using {\tt Pythia} and
list the cross sections in Table~\ref{smbkg}. These events are
passed through the same kinematic cuts used for the signal. The
background is quite small when compared with the signal
generated by doubly-charged Higgsino pairs. Indeed, as seen
from Fig.~\ref{pairprod}, the SM background nears the signal
only when $M_{\Dt}\sim 400\ {\rm GeV}$. This manifest dominance
of the signal makes this channel a highly promising one for an
efficient and clean disentanglement of LRSUSY effects.
\begin{center}
\bt[htb]
$$
\begin{array}{c} {\rm \bf \large SM~Background} \\
\begin{array}{|c|c|c|}\hline
\mbox{Final States} & 2\mu^-2e^+ +E\slash_T+{\rm X} & 2\mu^-e^++E\slash_T+{\rm X}\\
\hline \hline
 WW~\mbox{and}~ WZ &  0.02 ~{\rm fb} & 0.17 ~{\rm fb}\\
 t\bar{t}      &  0.14 ~{\rm fb} & 3.58 ~{\rm fb}\\ \hline
  {\rm Total}  &  0.16 ~{\rm fb} & 3.75 ~{\rm fb}\\ \hline \hline

\end{array}
\end{array}
$$
\caption{\sl\small The SM background cross sections for $2\mu^-2e^+ +E\slash_T$ and
$2\mu^-e^+ +E\slash_T$ in the final state at Tevatron. We have also put a $b$-jet veto
(assuming $b$-tagging efficiency to be 50\%) to suppress the background from the
production of $t\bar{t}$.}
\label{smbkg}
\et
\end{center}

\begin{figure}[htb]
\begin{center}
\includegraphics[height=3.0in,width=3.0in]{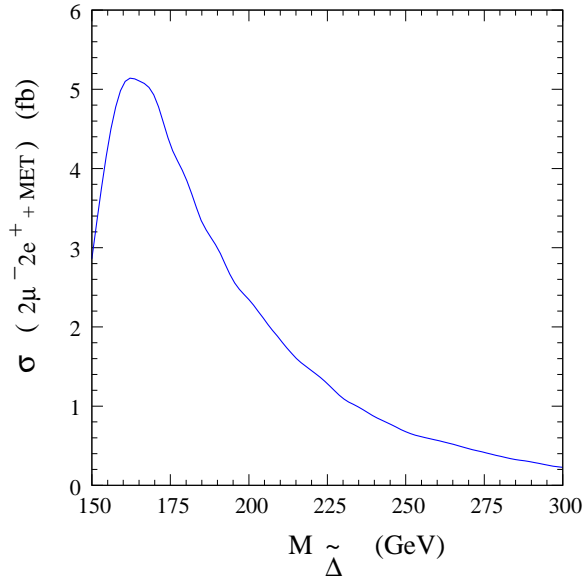}
\caption{\sl\small The signal cross section for the $2\mu^-2e^+ + E\slash_T$ final
state at the Tevatron (after selection cuts) in the LRSUSY model.}
\label{4lept}
\end{center}
\end{figure}

In Fig. \ref{4lept} we plot the total event cross sections
(after applying the kinematic cuts mentioned above). For the
four-lepton plus missing energy signal we have specifically
chosen the $2\mu^- + 2e^+ +E\slash_T$ final state. The total
cross section for a given final state is obtained by summing
over contributions of doubly-charged Higgsinos of either
chirality.
We find that the signal cross section is around $3\ {\rm fb}$
for $M_{\Dt}=150\gev$, and exhibits an increase for slightly
heavier $\Dt$ ($\simeq 175\gev$)  before the kinematic phase
space suppression takes over, as the mass of $\Dt$ is further
increased. This is quite expected, since the branching fraction
for the 3-body decay of $\Dt$ to $\ell\ell\N0_1$ depends on the
mass of $\Dt$ as shown in Fig. \ref{bratio}. The other decay
channel of $\Dt$ decay is $\ell\ell\N0_2$ which also gives two SSSF
leptons, with the $\N0_2$ decaying further to
$\ell^+\ell^-\N0_1$ or $q\bar{q}\N0_1$. This channel also
contributes to the signal, as the lepton pairs coming from the
$\N0_2$ decay are very soft (due the small mass
difference between $\N0_2$ and $\N0_1$), and might fail the
selection cuts. Even then the contributions are very small 
compared to the signal arising from $\Dt \to \ell\ell\N0_1$.
The total cross section lies between $2-3\ {\rm fb}$ for
$M_{\Dt}=200$ GeV but falls rapidly below $1\ {\rm fb}$ for
$M_{\Dt}
> 250\gev$ for the signal $2\mu^- 2e^+ +E\slash_T$. But if we
consider no charge identification and just take $2\mu 2e +
E\slash_T$, then the total cross section is
approximately twice the one from above. In principle, one can also work
with final states where one of the lepton flavors is $\tau$.
Then one needs to fold in the efficiencies for $\tau$
identification at the Tevatron to get the correct event rates.

Tevatron has already collected data samples with integrated 
luminosities
between $2-3\ {\rm fb}^{-1}$. With such a robust final state
one can expect excellent discovery potential for light
doubly-charged Higgsinos. Also, the fact that the production
rate for such a final state is mainly governed by the mass of
$\Dt$ should give us a very strong constraint on the mass of
the doubly-charged Higgsinos.

\begin{figure}[htb]
\begin{center}
\includegraphics[height=2.5in,width=2.5in]{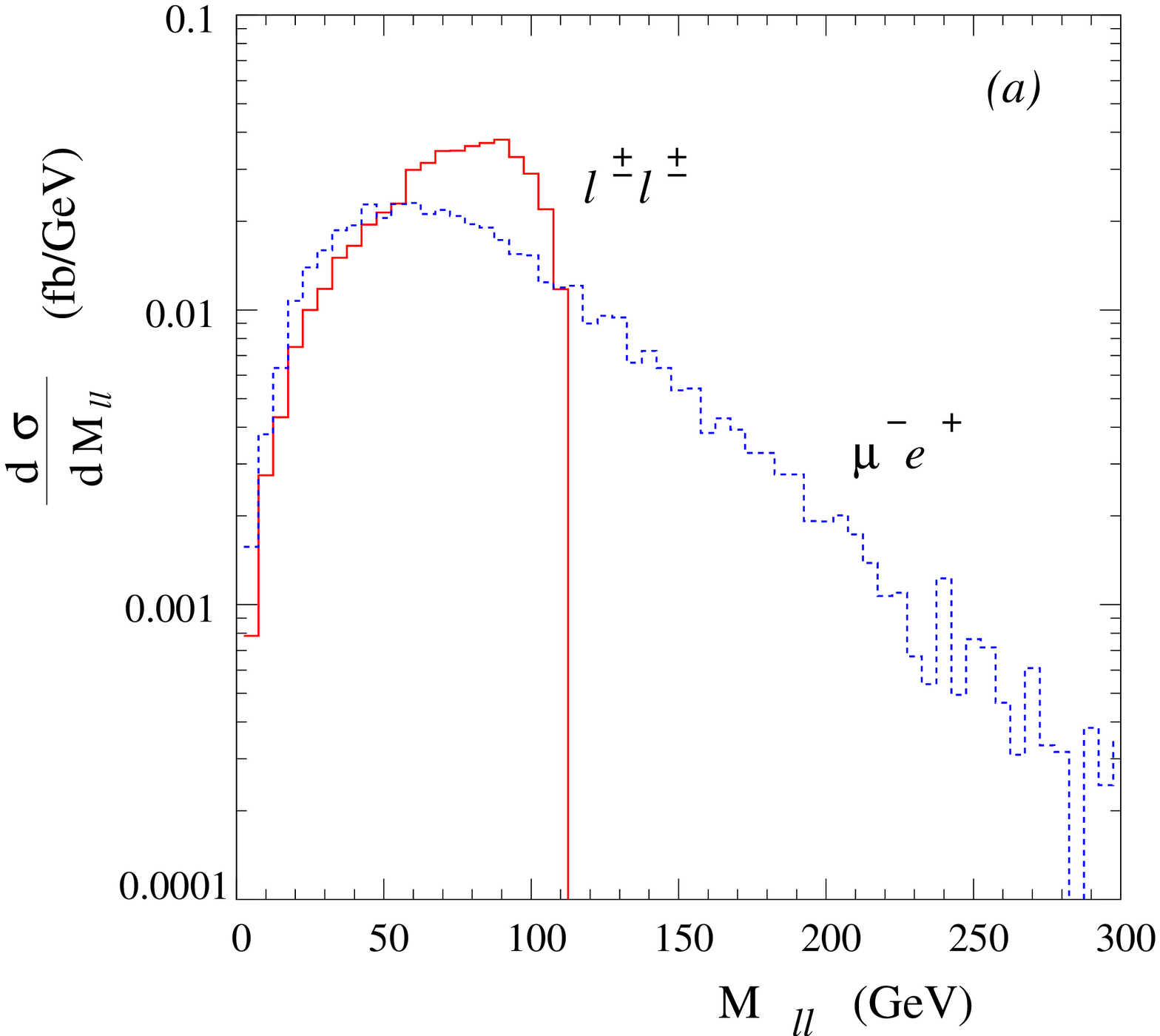}
\includegraphics[height=2.5in,width=2.5in]{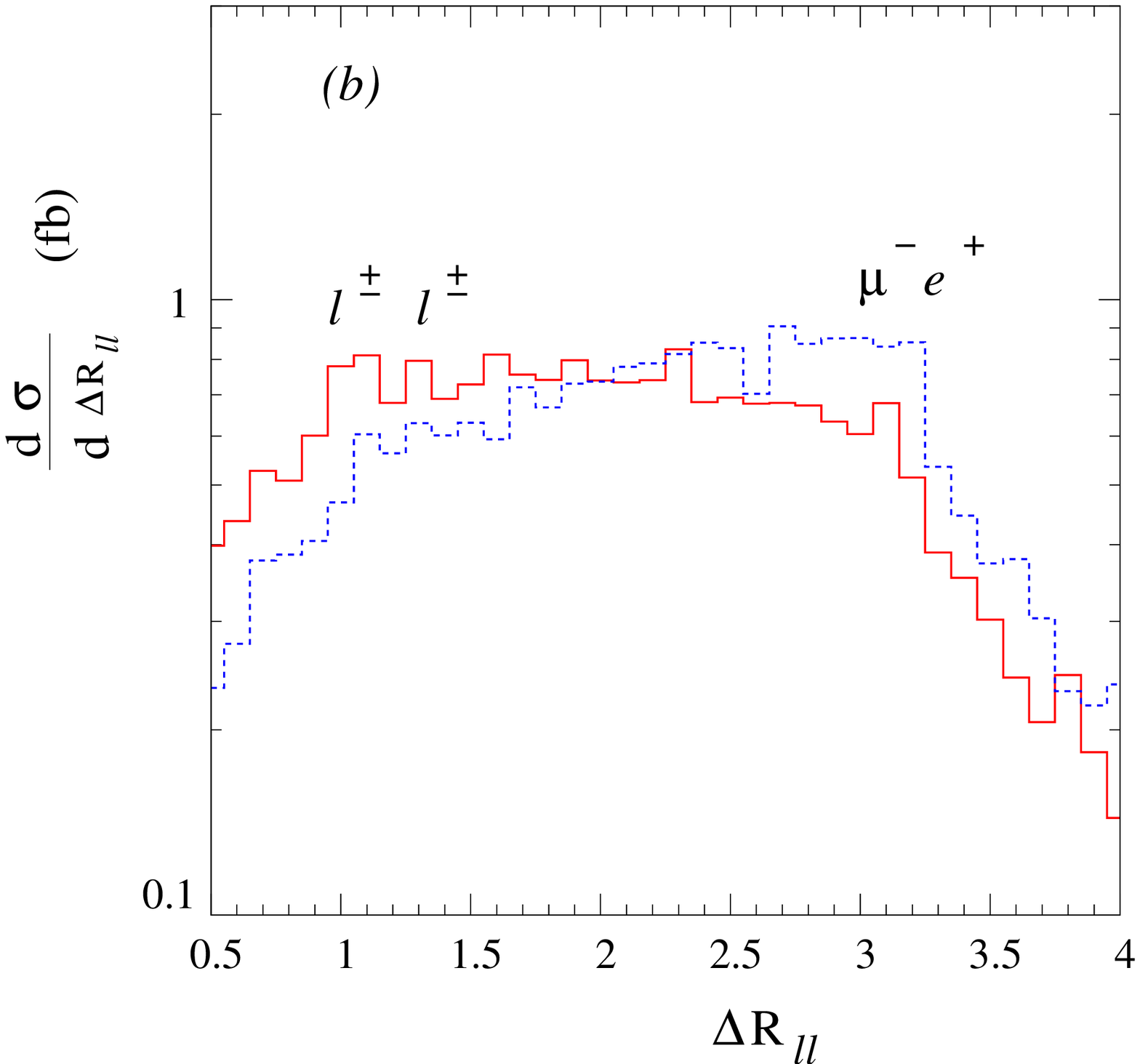}
\caption{\sl\small The differential cross section of the tetralepton signal
with respect to (a) the dilepton invariant mass and (b) $\D R$ of a 
dilepton pair,
for $M_{\Dt}=200\gev$. The binsize in panels (a) and (b) are 5 GeV 
and 0.1, respectively.}
\label{mlldll}
\end{center}
\end{figure}

In Fig.~\ref{mlldll} we choose $M_{\Dt}=200$ GeV for the {\it signal analysis} and
we show some characteristic kinematic features
unique to the model when compared with other new physics studies.
In Fig.~\ref{mlldll}(a) we plot the binwise distributions for
the invariant masses of the lepton pairs of the signal $2\mu^-
2e^+ + E\slash_T$. These plots manifestly show differences
between the SSSF and opposite-sign different-flavor (OSDF) lepton pairs with respect to their
invariant mass distributions. This is because the SSSF leptons
originate from the cascade decay of a single doubly-charged
Higgsino whereas OSDF lepton configurations are formed by two
isolated leptons, one originating from $\Dm$, the other from
$\Dp$.  For similar reasons, the SSSF lepton pairs exhibit a
sharp kinematic edge in their $M_{\ell\ell}$ distributions
whereas the OSDF lepton pairs do not. The reason is that SSSF
lepton pairs originate from the cascade decay of the same
$\Dt$. Since the dilepton invariant mass does not change under
boosts, this edge can be well-approximated by the formula (in
the rest frame of the decaying particle) \be
M_{\ell^\pm\ell^\pm}^{max} = \sqrt{M_{\Dt}^2 + M_{\N0_1}^2 - 2
M_{\Dt} E_{\N0_1}}\,\,, \label{eq:invmass} \ee where
$E_{\N0_1}$ is the energy of the LSP. This formula yields an
edge in the invariant mass distribution of the SSSF lepton
pairs at the bin around $M_{\ell^\pm\ell^\pm} = M_{\Dt} -
M_{\N0_1}$ (in agreement with the numerical values
corresponding to the {\bf SPA} point) in the case of the 3-body
decay of $\Dt$, as can be seen in Fig.~\ref{mlldll}(a). This
corresponds to the situation when the LSP is produced at rest
in the frame of $\Dt$.

A comparison of the SSSF and OSDF leptonic spectra qualifies to
be a viable probe of the doubly-charged Higgsinos. Firstly, the
edge in the SSSF dilepton invariant mass distribution yields a
clear hint of a $\D L=2$ interaction and a doubly-charged field
in the underlying model of `new physics'. The distributions of
the OSDF dileptons exhibit no such edge for the reasons
mentioned before. Secondly,  Fig.~\ref{mlldll}(b) manifestly
shows the differences in the binwise distribution of the
spatial resolution $\Delta R$ between the charged lepton pairs
for the cases indicated on the curves (with $l^\pm l^\pm$
standing for $\mu^-\mu^-$ or $e^+e^+$). The SSSF leptons have
distributions peaked at low values of $\D R$ (as they originate
from one single doubly-charged Higgsino $\Dt$) whereas the OSDF
leptons exhibit pronounced distributions at higher values of
$\D R$ (as they originate from different Higgsinos, one from
$\Dm$ the other from $\Dp$). To this end, SSSF leptons with
small spatial separation qualify to be a direct indication of
the doubly-charged Higgsinos in the spectrum. In addition to
the kinematic edge highlighted for the dilepton invariant mass,
this feature is a clear-cut signal of extended SUSY models as
it is absent in the MSSM or in any of its extensions that
contain only singly-charged fields.

\subsection{Associated productions of doubly-charged Higgsinos and Charginos}
\begin{figure}[htb]
\begin{center}
\includegraphics[height=3.0in,width=4.0in]{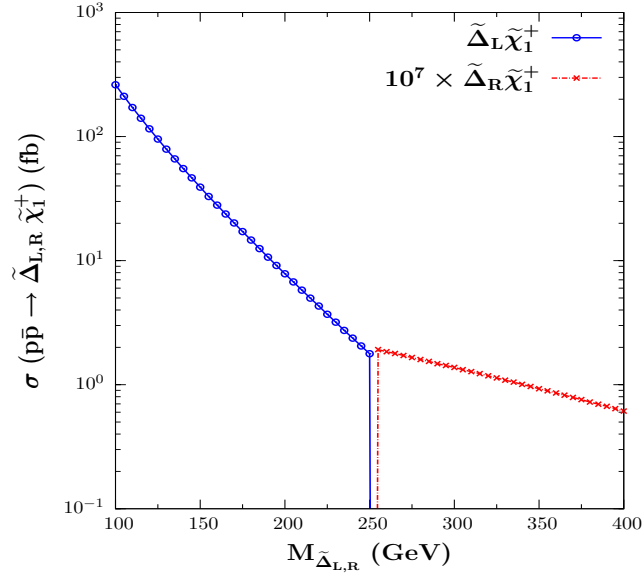}
\caption{\sl\small  The cross sections for productions of
$\Dt_{L,R}$ in association with $\tilde{\chi}_1^{\pm}$ at the Tevatron.
The model parameters are as in {\bf SPA} in Table \ref{susyin}, except that
$M_{\tilde \Delta^{--}} \equiv \mu_3$ is varied from $100\ {\rm GeV}$ up
to $400\gev$.}
\label{aprod}
\end{center}
\end{figure}

In this section we study productions and decays of doubly-charged Higgsinos
in association with the lightest chargino. The process under consideration
has the form
\be
p\, \bar{p} \longrightarrow \Dm\, \Cp_1 \longrightarrow
 \left(\ell_i^-\ell_i^-\right) +  \ell_j^+ + E\slash_T , \nonumber
\label{eq:singprod} \ee where $\ell_i$ is not necessarily
identical to $\ell_j$. This scattering process proceeds with
the $s$-channel $W_{L,R}$ exchange, and yields invariably a
trilepton signal, which has long been considered as a strong
indication of SUSY, in general \cite{Abbott:1997je}. In
Fig.~\ref{aprod} we plot the cross section for the associated
production of the $\Dt^{\pm\pm}$ which should give us a $3\ell
+ E\slash_T$ signal.
The cross section for singly-produced doubly-charged Higgsino
turns out to be small compared to the pair production cross
section. The production is completely dominated by the
left-chiral doubly-charged state for the model parameters ({\bf
SPA}) shown in Table \ref{susyin}. Indeed, the right-chiral
state is produced via the $s$-channel exchange of the
heavy $W_R$ boson, and it is thus strongly suppressed. There is
also additional suppression due to the composition of the
lightest chargino, which affects the coupling of
$W_R^\mu-\Cp_1-\Dm_R$. The couplings of the chargino states to
weak gauge bosons and the doubly-charged Higgsino are

$\displaystyle
\bullet ~~W^{\mu}_L ~\tilde{\chi}_k^+ ~\tilde \Delta_L^{--}:
~~~~ ig_L\gamma^{\mu}(V^{\star}_{k5}P_L+U_{k5}P_R)$,

$\displaystyle \bullet ~~W^{\mu}_R ~\tilde{\chi}_k^+ ~\tilde
\Delta_R^{--}: ~~~~ ig_R\gamma^{\mu}(V^{\star}_{k6}P_L+U_{k6}P_R)$,

\noindent where $U$ and $V$ are the matrices which diagonalize
the chargino mass matrix. The production cross section becomes
negligible around $M_{\widetilde{\Delta}}=250\gev$, at which the
composition of the lightest chargino changes abruptly. More
precisely, the lightest chargino is dominantly triplet Higgsino
for $\mu_3 \le 250\gev$, for which the coupling $W-\Cp_1-\Dt_L$
is maximal. For $\mu_3 > M_L$,   there exists a
non-negligible gaugino contamination in the lightest chargino,
and the $W-\Cp_1-\Dt_L$   depends on the left-triplet
VEV, which suppresses it below any measurable level. This also
explains the fact that the associated single production of
$\Dt$ with the lightest chargino will identically vanish as
soon its the gaugino component starts becoming non-negligible.
Thus, the $3\ell+ E\slash_T$ signal with a pair of SSSF leptons
depends strongly  on the composition of the chargino.

Consequently, it suffices to use only the left-chirality
doubly-charged Higgsino production in association with the
lightest chargino $\tilde{\chi}_1^{+}$. Nevertheless, we note
that for the choice of parameters in ({\bf SPA}), as the
composition of the chargino changes around $\mu_3 \ge 250\gev$, 
the associated production of $\Cp_2\Dm_L$ begins to dominate.
Fig.~\ref{aprod} shows that, for the entire {\bf SPA} parameter
space with varying $\mu_3$, the left-chirality doubly-charged
Higgsino produced in association with the lightest chargino
yields a reasonable cross section for small Higgsino masses,
and yields approximately $2\ {\rm fb}$ cross section for
$\Dt^{\pm\pm}$ as heavy as  $M_{\Dt} \sim 250\gev$. For
comparison, we also include the cross section for the
right-handed Higgsino, which is small, even when $\mu_3 >
250\gev$ ( because of the strong $s$-channel suppression
and nature of the composition of the lightest chargino, as
pointed out earlier). One notes here that, since the chargino
couplings to $\Dt_{L/R}$ depend on the entries in the mixing
matrices of charginos, the input parameters in Table
\ref{susyin} play a crucial role in determining the production
cross section. The cross section for $p\, \bar{p} \to \Dm_L\,
\Cp_1$ is around $8\ {\rm fb}$ for {\bf SPA} parameter set. As
in pair-production mode, the $\Dm$ decays again into a pair of
SSSF leptons and the LSP following (\ref{eq:decays}). However,
the chargino, too, exhibits a three-body
decay through off-shell sleptons with almost $100\%$ branching
into the three-body final state of neutrino, lepton and the LSP for
{\bf SPA} parameter set with $M_{\Dt} \ge 175\gev$. For
lower values of $\mu_3=M_{\Dt}$, however, the mode
$\bar{\nu}_\ell \ell^+ \N0_2$ competes. The above decays yield
a $3\ell+E\slash_T$ final state, where the missing transverse
energy arises from the undetected LSP and the neutrino. For the
benchmark point {\bf SPA}, the signal acquires all the contribution
from the left-chirality state production.

The single $\Dm$ production gives rise to a trilepton signal at
the Tevatron experiments. In the numerical analysis, following
the same notation and same kinematic cuts as in the previous
subsection (for the three leading charged leptons), we
illustrate the case where $\ell_i=\mu$ and $\ell_j=e$. Thus, we
know that the $e^+$ always comes from the chargino while the
same-sign muons originate from the doubly-charged Higgsino.
\begin{figure}[htb]
\begin{center}
\includegraphics[height=3.0in,width=3.0in]{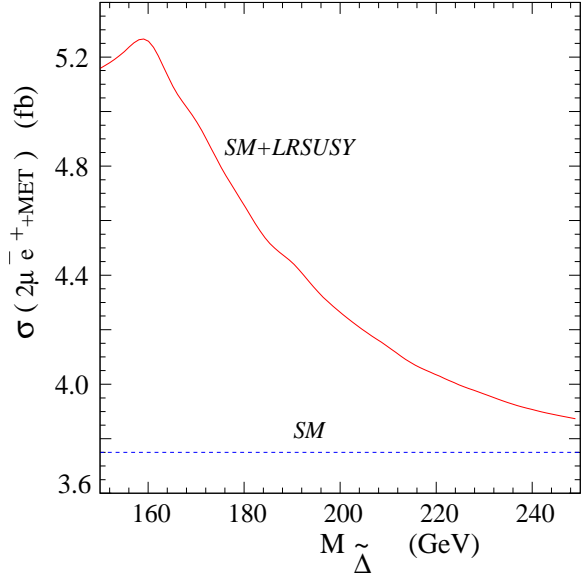}
\caption{\sl\small The cross section for $2\mu^-e^+ + E\slash_T$ final
state at the Tevatron (after selection cuts) at the Tevatron.}
\label{3lept}
\end{center}
\end{figure}
In parallel to the analysis of $4\ell+E\slash_T$ signal in
previous subsection, the characteristic kinematic features
remain very similar to the ones that have already been covered
by discussions in the previous subsection. In particular, the
distributions of the SSSF leptons are quite similar to the ones
for the $4\ell+E\slash_T$ signal. This is actually expected
since SSSF leptons are exclusively generated by the decays of
the doubly-charged Higgsino, a common feature for both
tetralepton and trilepton final states.

However, we note that the SM background is substantial for the
trilepton final state at the Tevatron energies. It stands at
$3.75\ {\rm fb}$ level with the most dominant source being the
$t\bar{t}$ production. We list the major backgrounds in
Table~\ref{smbkg}. We have plotted the signal+SM cross section
in Fig.~\ref{3lept} by considering specifically $2\ell^-_i +
\ell^+_j + E\slash_T$ signal with $\ell_i = \mu$ and $\ell_j =
e$. It shows that an integrated luminosity of $10\ {\rm
fb}^{-1}$ does not give a $3\sigma$ signal over the background
in the trilepton channel for {\bf SPA}. For an integrated
luminosity of $16\ {\rm fb}^{-1}$ the signal reach at $2\sigma$
is for $M_{\Dt} \lsim 178\gev$ while the reach is only extended at
most to $M_{\Dt} \lsim 154-162\gev$ for a $3\sigma$ signal.
The SM background can be reduced further to $1.52\ {\rm
fb}$ by choosing a set of larger cuts for $p_T$, namely
\{20,15,10\} GeV (taken in the same order as
before).
This helps in extending the reach for a $2\sigma$ signal 
to doubly-charged Higgsino masses of $M_{\Dt} \lsim 190\gev$
while the $3\sigma$ reach stands at $M_{\Dt} \lsim 173\gev$.

Nonetheless, the pair production of doubly-charged Higgsinos
for the same representative point {\bf SPA} gives at least 2
(8) events for the tetralepton signal for $M_{\Dt}=280\gev$,
for an integrated luminosity of $4 (16)\ {\rm fb}^{-1}$ where 0
(3) SM events are expected, giving thus a clear discovery
signal for the doubly charged exotica.

\section{Discussions and Conclusions}

We have studied the Tevatron signals of doubly-charged
Higgsinos present in the left-right symmetric SUSY models. The
doubly-charged Higgsinos in the spectrum are a
characteristic feature of LRSUSY, which can directly and
unambiguously distinguish the model from the MSSM (and its
various extensions like NMSSM and U(1)$^{\prime}$ models) via
generic leptonic events observed in collider experiments.
We have given a detailed account of the leptonic signals
originating from production-and-decay of $(i)$ doubly-charged
Higgsino pairs (in Sec. 3.1) and of $(ii)$ single
doubly-charged Higgsino plus chargino (in Sec. 3.2). For the
production mode $(i)$,  the leptonic final state invariably
involves $\left(\ell_i^-\ell_i^-\right) +
\left(\ell_j^{+}\ell_j^{+}\right) + E\slash_T$, that is, a pair
of SSSF dileptons plus missing energy acquired by the LSP,
$\tilde{\chi}_1^0$. On the other hand, for the production mode
$(ii)$ the leptonic final state is composed of $\left(\ell_i^-
\ell_i^-\right) + \ell_j^{+} + E\slash_T$, that is, a trilepton
signal.

The simulation study of the pair-production process,
characterized by tetralepton final states with SSSF or OSDF
structures, provides a unique opportunity to track down the
presence of doubly-charged fermions in the model of 'new
physics' at the TeV domain. Especially noticeable are the
distributions of the SSSF  leptons,  which can firmly
establish if there exists a doubly-charged fermion that decays
into SSSF leptons plus missing energy (by using the dilepton
invariant mass and their spatial proximity). Also important is
the fact that the SSSF and OSDF lepton signals well dominate
over the SM signal up to relatively large doubly-charged
Higgsino masses.

The simulation study of the production of single doubly-charged
Higgsino in association with chargino, which yields the
celebrated trilepton signal, also proves to be an important
signature of doubly-charged fermions of `new physics'.
Nevertheless, the signal dominates over the SM background only
for low Higgsino masses, and one cannot extract the information
about doubly-charged fermions as confidently as in the
tetralepton signal, above.

The search programme in this work can be extended to a
multitude of `new physics' models at both qualitative and
quantitative level. Generically, however, the LRSUSY, whose
spectrum consists of doubly-charged Higgs fermions, stands
fundamentally different than the rest as it possesses SSSF
proximate dileptons at the final state. In this sense, the
analysis in this work shows that, under minimal assumptions
about the model and its parameter space ${\bf SPA}$, as far as
detector acceptance permits, the fingerprints of light
doubly-charged Higgsinos can be searched in the $4\ {\rm
fb}^{-1}$ of data collected by the CDF and D0 experiments. This
requires performing the requisite `event mining' in Tevatron
data exclusively for tetralepton final states with SSSF
structure. Such exotic structures, if discovered, would provide a spectacular signal for physics beyond SM and MSSM.

\section{Acknowledgements}
The work of D.D. was supported by the Turkish Academy of
Sciences via GEBIP grant, by the Turkish Atomic Energy Agency
via CERN-CMS grant, and by IZTECH via 2009 BAP grant. 
The work of M.F. and I.T. was supported in part by NSERC of
Canada under the Grant No. SAP01105354. D.K.G. acknowledges the partial support from the Department of Science and Technology, India under grant SR/S2/HEP-12/2006. K.H. and S.K.R. gratefully acknowledge support from the Academy of
Finland (Project No. 115032). We thank D. Choudhury and B. Mukhopadhyaya for discussions. D.K.G. would like to acknowledge the hospitality provided by the Helsinki Institute of Physics. D.K.G., K.H. and S.K.R. thank Nordita program {\it TeV scale physics and dark matter} for an inspiring atmosphere.


\end{document}